\documentclass{ws-procs975x65}

\begin{document}

\title{Energy landscapes, scale-free networks and Apollonian packings}

\author{Jonathan~P.~K. Doye and Claire P. Massen}

\address{University Chemical Laboratory, Lensfield Road, Cambridge CB2 1EW, 
United Kingdom}

\maketitle

\abstracts{We review recent results on the topological properties
of two spatial scale-free networks, the inherent structure and 
Apollonian networks. The similarities between these two types of network
suggest an explanation for the scale-free character of the inherent
structure networks. Namely, that the energy landscape can be viewed as
a fractal packing of basins of attraction.}

\section{Introduction}

The potential energy as a function of the coordinates of all the atoms in a 
system defines a multi-dimensional surface that is commonly known as an 
energy landscape.\cite{Wales03} 
Characterizing such energy landscapes has become an 
increasingly popular approach to study the behaviour of complex systems,
such as the folding of a protein\cite{Bryngelson95} 
or the properties of supercooled liquids.\cite{Still95,Debenedetti01}
The aim is to answer such questions as, 
what features of the energy landscape differentiate those polypeptides
that are able to fold from those that get stuck in the morass of possible
conformations, or those liquids that show super-Arrhenius dynamics
(`fragile' liquids) from those that are merely Arrhenius (`strong' liquids).

\begin{figure}[t] 
\centerline{\epsfxsize=5.0in\epsfbox{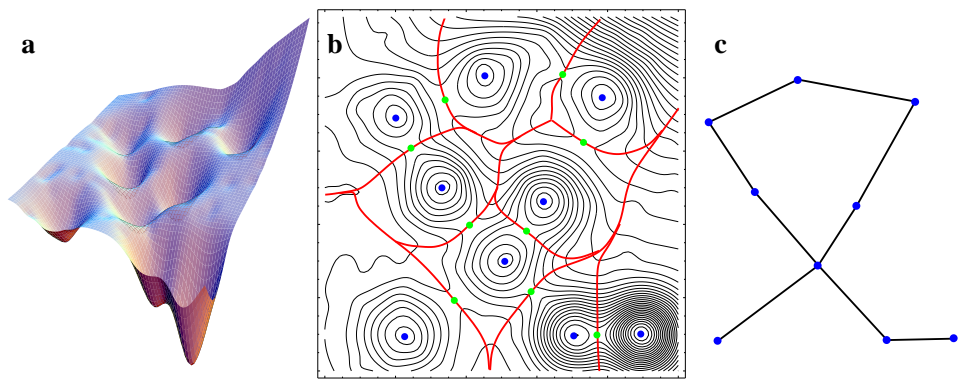}}   
\caption{(a) A model two-dimensional potential energy surface, 
(b) the contour plot of this surface showing the
`inherent structure' division of 
the energy landscape into basins of attraction
(the minima and transition states are represented by points and 
the basin boundaries by the thick lines), and 
(c) the representation of the landscape as a network.
\label{fig:PES}}
\end{figure}

Such approaches have to be able to cope with the complexity 
of the potential energy landscape---for example, 
the number of minima is typically an exponential function
of the number of atoms.\cite{Still99} 
One such approach is the
inherent structure mapping pioneered by Stillinger and 
coworkers.\cite{StillW84a} In this mapping each point in configuration space is
associated with the minimum obtained by following the steepest-descent pathway
from that point.  Thus, configuration space is partitioned into a set of 
basins of attraction surrounding the potential energy minima, 
as illustrated in Fig. \ref{fig:PES}.

One of the original aims of this approach was to remove
the vibrational motion from configurations generated in simulations 
of liquids to give a clearer picture of the underlying `inherent structure',
hence the common name for the mapping.
Of more interest to us is that it breaks the energy landscape down into 
more manageable chunks, whose properties can be more easily established 
and understood.
As an example of the utility of this approach, 
the classical partition function can be expressed as an
integral over the whole of configuration space, but performing
this integral (except numerically through say Monte Carlo) is nigh
impossible, because of the complexity of the potential energy landscape.
However, if this integral is divided up into separate integrals over
each basin of attraction, analytical approximations to these individual 
integrals can easily be obtained by assuming that the basins can be modelled as 
a harmonic well surrounding the minimum at the centre of the basin.
Calculation of an approximate partition function, then just reduces to 
a characterization of the properties of the potential
energy minima and their associated basins.\cite{Wales93a}
As well as providing insights into the contributions of different regions
of the energy landscape to the thermodynamics, quantitative accuracy can 
be obtained when account is taken of the anharmonicity of the 
basins.\cite{Calvo01e}

Similarly, an energy landscape perspective on the dynamics can be formulated
in terms of the transitions between the basins of attraction.
Except at sufficiently high temperature, a trajectory of a system 
can be represented as a series of episodes of vibrational motion 
within a basin, punctuated by occasional hopping between basins 
along a transition state valley.\cite{Goldstein69} 
In a coarse-grained view that ignores
the vibrational motion, the dynamics is a walk on a network
where the nodes correspond to minima and there are edges between minima
that are directly connected by a transition state.\footnote{By 
a transition state we mean a 
stationary point on the potential energy landscape that has one eigendirection
with negative curvature. The steepest-descent pathways from the transition
state parallel and anti-parallel to this Hessian eigenvector then provide
a unique definition of the two minima connected by this transition state.}
An example of such an `inherent structure' network is also 
illustrated in Fig.\ \ref{fig:PES}.

Although there has been much work characterizing energy landscapes
with the aim of gaining insights into particular systems, some of the 
fundamental properties of such landscapes, particularly those related
to their global structure and organization, have received relatively 
little attention. 
For example, what is the nature of the division of the energy landscape
into basins of attraction and does the inherent structure network have a 
universal form? 
In this chapter, we will be reviewing recent results that address exactly 
these questions. 
The system that we will be analysing is a series of small 
Lennard-Jones (LJ) clusters for which 
the complete inherent structure network can be found.

Another approach to understanding the properties of complex systems 
that has received much attention recently is through an analysis 
of the system in terms of networks.\cite{Newman03a,Albert02}
The systems analysed in this way have spanned an
impressive range of fields, including astrophysics,\cite{Hughes03}
geophysics,\cite{Baiesi04} information technology,\cite{Albert99}
biochemistry,\cite{Jeong00,Jeong01}
ecology,\cite{Dunne02} and sociology.\cite{Liljeros01}
Initially, the focus was on relatively basic topological properties of
these networks, such as the average separation between nodes and
the clustering coefficient to test whether they behaved like the
Watts-Strogatz small-world networks,\cite{Watts98} or
the degree distribution\footnote{In network parlance, the degree $k$
is the number of connections to a node.} 
to see if they could be classified as scale-free
networks.\cite{Barabasi99}

To summarize our recent results 
we found that the inherent structure networks associated 
with the LJ clusters behaved as fairly typical scale-free 
networks.\cite{Doye02c,Doye05b,Massen05a}
However, the origins of most scale-free networks can be explained in terms
of network growth models, where there is preferential attachment to 
nodes with high degree during network growth.\cite{Barabasi99}  
By contrast, the inherent structure networks are static. They are 
determined just by the potential describing the interatomic interactions
and the number of atoms in the system. So,
why are they scale free?

\begin{figure}[t] 
\centerline{\epsfxsize=5.0in\epsfbox{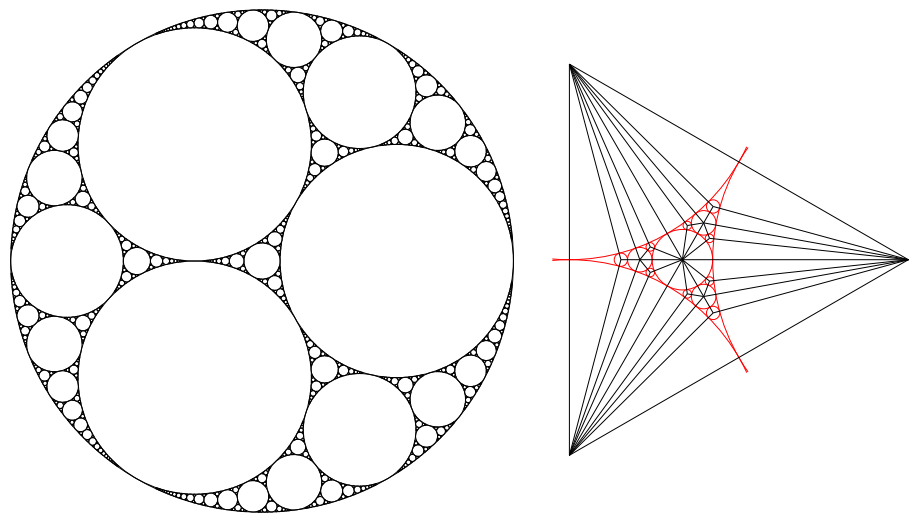}}   
\caption{The Apollonian packing of a circle, and the corresponding
network for the central interstice between the initial disks after
three generations of disks have been added.
\label{fig:Apollo}}
\end{figure}

One of the important features of the inherent structure networks is 
their embedding in configuration space.
There have been a number of model spatial scale-free networks 
proposed,\cite{Rozenfeld02,Warren02,benAvraham03,Herrmann03} but
the ones on which we wish to focus are Apollonian 
networks.\cite{Andrade04,Doye05a} These networks are associated with
Apollonian packings, an example of which is given in Fig.\ \ref{fig:Apollo}.
To generate such a packing, one starts with a set of touching disks 
(or hyperspheres if one is interested in higher-dimensional packings)
and then to each interstice in the packing, new disks are added 
that touch each disk surrounding the interstice.
At each subsequent generation the same procedure of adding disks
to the remaining interstices is applied. 
The complete space-filling packing is obtained by repeating
this process {\it ad infinitum}.
The Apollonian network is then the contact network between adjacent 
disks (Fig.\ \ref{fig:Apollo}).

One of the reasons that the Apollonian network provides a 
useful comparison to the inherent structure networks is that
spatial regions (the disks) are automatically associated with
each node in the network, which is somewhat similar to the association
of the basins of attraction with the minima on an energy landscape. 
Furthermore, in both networks 
edges are based on contacts between those spatial regions that are adjacent.
As a consequence, for two-dimensional examples, both types of network are 
planar, that is, they can be represented on a plane without any edges 
crossing.\cite{Aste04}
This feature contrasts with the other model spatial scale-free 
networks.\cite{Rozenfeld02,Warren02,benAvraham03,Herrmann03} 
Therefore, in this chapter we will be comparing the
properties of the inherent structure and Apollonian networks.

\section{Comparing Apollonian and inherent structure networks}

\begin{figure}[t] 
\centerline{\epsfxsize=5.0in\epsfbox{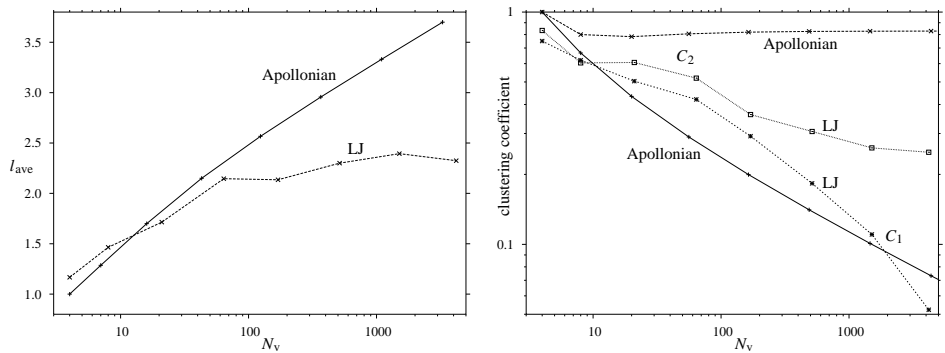}}   
\caption{The scaling of the average separation between nodes and 
the clustering coefficient with network size for LJ clusters with 7
to 14 atoms, and two-dimensional Apollonian networks with increasing numbers
of generations.
\label{fig:smallworld}}
\end{figure}

We were able to obtain the complete inherent structure networks for 
all LJ clusters with up to 14 atoms. For the largest cluster, 
the network had 4196 nodes and $87\,219$ edges. By contrast, the Apollonian
networks have an infinite number of nodes. Therefore, to allow
a comparison we consider finite Apollonian networks obtained
by only considering 
the first $t$ generations of disks.
The comparison we make is usually between the LJ$_{14}$ network and a
two-dimensional Apollonian network with a similar number of nodes 
(in fact with $t=7$ and 4376 nodes and $13\,122$ edges).
One could argue that it would be more appropriate to compare to an Apollonian 
network with the same spatial dimension. However, the properties of the
Apollonian networks are very similar irrespective of dimension, so 
we chose to use the two-dimensional example simply because the properties
of this case have been most comprehensively worked out.

To study the size dependence of the network properties, as in Fig.\ 
\ref{fig:smallworld} we have to make a further choice. 
For the inherent structure networks we follow clusters with an 
increasing number of atoms, and hence an increasing dimension
of configuration space.
Again it could be argued that we should be comparing
to an Apolonian network of fixed number of generations, but increasing dimension,
but the useful feature of examining a network of fixed dimension and
increasing $t$ instead is that the variable $t$ behaves in a somewhat similar way 
to the number of atoms. For example, the number of nodes is an 
exponential function of $t$, whereas it only increases polynomially
with the dimension of the system.\cite{Doye05a}  As already mentioned, the 
number of minima is an exponential function of the number of 
atoms.

\begin{figure}[t] 
\centerline{\epsfxsize=3.0in\epsfbox{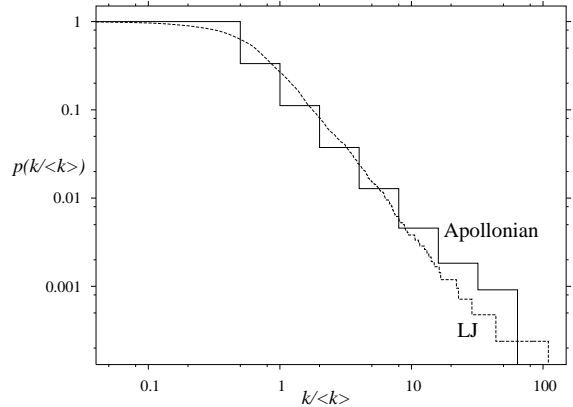}}   
\caption{The cumulative degree distributions for the inherent structure and Apollonian
networks.
\label{fig:degree}}
\end{figure}

From Fig.\ \ref{fig:smallworld}, one can see that both types of networks
have small-world properties. Firstly, for both networks the average separation
between nodes scales no more than logarithmically with system size, as for 
a random graph. The stronger sub-logarithmic behaviour for the inherent 
structure networks is because the average degree increases with network size
(the random graph result is in fact $l_{\rm ave}=\log N_v/\log\langle k\rangle$)
whereas it is approximately constant for the Apollonian networks.
The increase in $\langle k\rangle$ is simply because the ratio of the
number of transition states to minima on a potential energy landscape
is a linear function of the number of atoms.\cite{Doye02a}
Secondly, the clustering coefficient, one measure of the local ordering
within a network, has values that are significantly larger than for 
a random network. The size dependence of this property depends on how it
is defined. If it is as the probability that any pair of nodes with a common
neighbour are themselves connected ($C_1$) then it decreases quite rapidly 
with size.
The second definition ($C_2$) is as the average of the local clustering 
coefficient, where the latter is defined as the probability 
that the neighbours of a particular node 
are themselves connected. The second definition gives more weight to the 
low-degree nodes that, as we shall see later, have a higher local 
clustering coefficient. That $C_2$ tends to a constant value 
for the Apollonian network, rather than decaying weakly as for the inherent
structure networks, reflects the stronger degree dependence of the local 
clustering coefficient.

Both networks also have a power-law tail to their degree distribution,
and so are scale-free networks. The exponent is slightly larger 
for the inherent structure networks (2.78 compared to 2.59).
This heterogeneous degree distribution is easier to understand for the
Apollonian network, and reflects the fractal nature of the
packings.\cite{Mandelbrot} At each stage in the generation of the network, the 
degrees of the nodes double, i.e.\ new nodes preferentially connect
to those with higher degree, and so the highest degree nodes correspond to 
those that are `oldest' and have larger associated disks. 

\begin{figure}[t] 
\centerline{\epsfxsize=3.0in\epsfbox{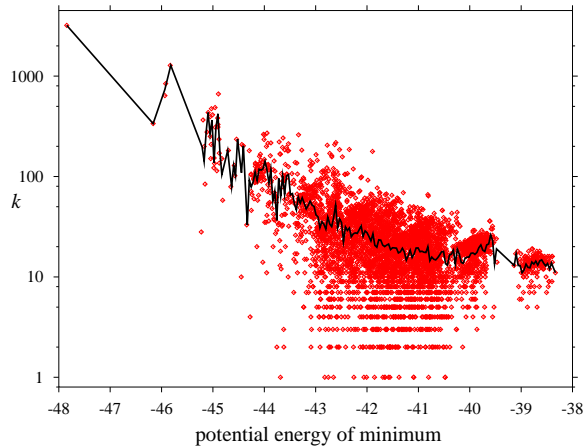}}   
\caption{The dependence of the degree of a node on the potential energy
of the corresponding minimum for
LJ$_{14}$.
The data points are
for each individual minimum and the solid line is a binned average.
\label{fig:kvE}}
\end{figure}

For the inherent structure networks, the high-degree nodes correspond to 
minima with low potential energy (Fig.\ \ref{fig:kvE}).
Our rationale for this correlation between degree and potential energy
is that the lower-energy minima have larger basin areas,\cite{Doye98e} 
and hence longer basin boundaries with more transition states on them.
The scale-free character of these networks must reflect the hierarchical
packing of these basins with larger basins surrounded by smaller basins, which
in turn are surrounded by smaller basins, and so on, 
in a manner somewhat similar to the Apollonian packing. Thus, 
the comparison of the inherent structure and Apollonian networks can 
provide some check of the plausibility of this potential origin of the
scale-free behaviour of the inherent structure networks.

\begin{figure}[t] 
\centerline{\epsfxsize=5.0in\epsfbox{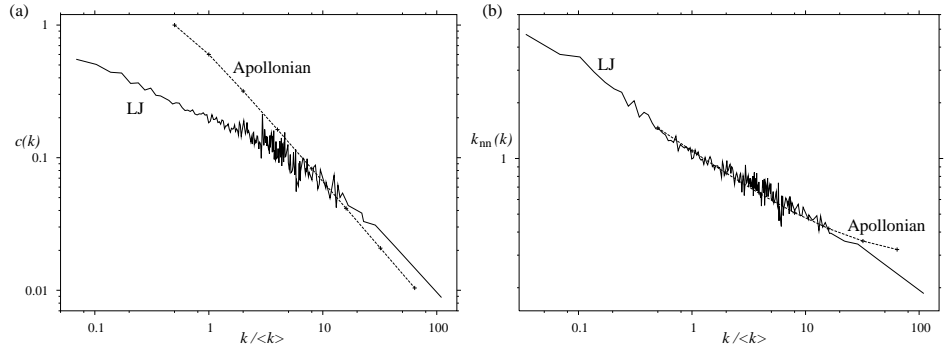}}   
\caption{The degree dependence of (a) the local clustering coefficient and
(b) $k_{\rm nn}$, the average degree of the neighbours of a node
for the inherent structure and Apollonian networks. 
Both lines represent the average values for a given $k$.
\label{fig:kdepend}}
\end{figure}

Figure \ref{fig:kdepend} shows that the two types of networks also behave 
similarly when we look at more detailed properties of the networks. Both have a 
local clustering coefficient that decreases strongly with increasing degree.
For the Apollonian network, it is actually inversely proportional to the 
degree\cite{Doye05a} --- a feature that has been previously seen for other 
deterministic scale-free networks\cite{Dorogovtsev02,Ravasz03,Comellas04} and 
that has been interpreted in terms of a 
hierarchical structure to the network\cite{Ravasz03,Barabasi04} ---
whereas for the inherent structure networks the degree dependence is 
somewhat reduced at small $k$.
This similar behaviour partly reflects the common spatial character 
of the networks. 
The smaller low-degree nodes have a more localized character and so their
neighbours are more likely to be connected, whereas the larger high-degree
nodes can connect nodes that are spatially distant from each other and so 
are less likely to be connected.

The behaviour of $c(k)$ also partly reflects the correlations\cite{Soffer04} evident in 
Fig. \ref{fig:kdepend}(b). 
Both networks are disassortative, 
that is nodes are more likely to be connected to nodes with dissimilar degree.
By contrast, for an uncorrelated network, $k_{nn}(k)$ would be 
independent of degree. However, it is well known that disassortativity can 
arise for networks, as here, in which multiple edges and self-connections are 
not present.\cite{Park03}
Indeed, for the inherent structure networks $k_{nn}(k)$ 
for a random network with the same degree distribution looks almost 
identical.\cite{Doye05b} An additional source of disassortativity is present in the 
Apollonian networks, because, except for the initial disks, there are no 
edges whatsoever between nodes with the same degree; disks created in the
same generation all go in separate interstices in the structure and so cannot
be connected.
Therefore, that $k_{nn}(k)$ for the two types of networks
follow each other quite so closely is probably somewhat accidental.

\begin{figure}[t] 
\centerline{\epsfxsize=3.0in\epsfbox{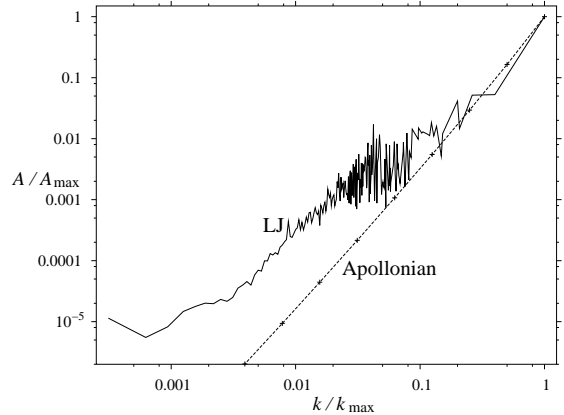}}   
\caption{The degree dependence of the basin areas for the LJ$_{14}$
energy landscape and disk areas for the Apollonian packing.
Both lines represent the average values for a given $k$.
\label{fig:Avk}}
\end{figure}

The behaviour seen for most of the network properties discussed 
so far is fairly common for scale-free networks. Therefore, a better
test of the applicability of the Apollonian analogy to the energy landscape
is to examine the spatial properties of the two systems directly. 
For the inherent structure networks, in agreement with the suggestion
made earlier, there is a strong correlation between the degree of a node
and the hyperarea of the basin of attraction that is similar to the
degree dependence of the disk area seen for the Apollonian networks 
(Fig.\ \ref{fig:Avk}). This result therefore implies that there is also 
a strong dependence of the basin area on the energy of a minimum with the
low-energy minima having the largest basins. It also provides
strong evidence that the scale-free topology of the inherent structure
networks reflects the heterogeneous distribution of basin areas.

\begin{figure}[t] 
\centerline{\epsfxsize=3.0in\epsfbox{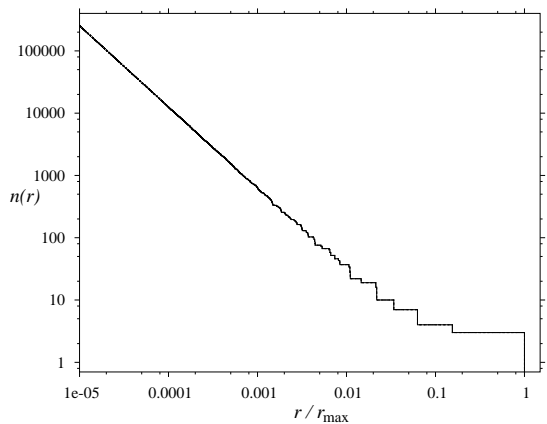}}   
\caption{The cumulative distribution for disks with radius greater than $r$ in a
two-dimensional Apollonian packing.
\label{fig:pr}}
\end{figure}

The distribution of disk areas for the Apollonian packing reflects its
fractal character.\cite{Mandelbrot} 
It is in fact a power-law\cite{Melzak69} with an exponent that
depends upon the fractal dimension of the packing,\cite{Manna91} as
illustrated in Fig.\ \ref{fig:pr}
For high-dimensional
packings this exponent tends to $-2$.\cite{Doye05a}
Preliminary results suggest that there is a similar power-law distribution for
the hyperareas of the basins of attraction on an energy landscape, 
confirming the deep similarity between these two types  of system,
and suggesting that configuration space is covered by a fractal packing 
of the basins of attraction.

\section{Conclusion}

In this chapter we have looked at some of the fundamental organizing principles
of complex multi-dimensional energy landscapes. By viewing 
the landscapes as a network of minima that are linked by transition states, 
we have found that the topology of this network is scale-free. 
Unlike most scale-free networks, the origin of this topology must be static.
We believe that it is driven by a very heterogeneous size distribution for 
the basins of attraction associated with the minima, with the large basins
having many connections. In this paper, we have explored 
whether space-filling packings of disks and hyperspheres, such as the
Apollonian packings, and their associated contact networks 
can provide a good model of how the energy landscape is organized. 
We have shown that these systems share a deep similarity both
in the topological properties of the networks and the spatial properties
of the packings. In fact, our results suggest that the energy 
landscape can be viewed as a fractal packing of basins of attraction.
Although this conclusion can provide an explanation for the scale-free topology
of the inherent structure network, it itself demands an explanation. Why are the
basins of attraction organized in this fractal manner? We will explore this in 
future work.


\end{document}